\documentclass[12pt]{iopart}
\usepackage{iopams}
\begin{document}
\title[Laplace transformations and sine-Gordon type integrable PDE]{Laplace transformations and sine-Gordon type integrable PDE}

\author{I T Habibullin$^{1}$, K I Faizulina$^{2}$, A R Khakimova$^{1}$}

\address{$^1$Institute of Mathematics, Ufa Federal Research Centre, Russian Academy of Sciences,
112, Chernyshevsky Street, Ufa 450008, Russian Federation}
\address{$^2$Ufa State Petroleum Technological University, 1, Kosmonavtov street, Ufa 450064, Russian Federation} 
\eads{\mailto{habibullinismagil@gmail.com}, \mailto{cherkira@mail.ru}, \mailto{aigul.khakimova@mail.ru}}

\begin{abstract}
It is well known that the Laplace cascade method is an effective tool for constructing solutions to linear equations of hyperbolic type, as well as nonlinear equations of the Liouville type. The connection between the Laplace method and soliton equations of hyperbolic type remains less studied. The article shows that the Laplace cascade also has important applications in the theory of hyperbolic equations of the soliton type. Laplace's method provides a simple way to construct such fundamental objects related to integrability theory as the recursion operator, the Lax pair and Dubrovin-type equations, allowing one to find algebro-geometric solutions. As an application of this approach, previously unknown recursion operators and Lax pairs are found for two nonlinear integrable equations of the sine-Gordon type.
\end{abstract}

\maketitle

\eqnobysec

\section{Introduction}

More than two centuries ago Laplace discovered an exceptionally effective method for solving linear second order hyperbolic type equations with variable coefficients \cite{Laplace}. His discovery had a huge impact on the further development of the theory of non-linear partial differential equations. Laplace's cascade method also finds important applications in the modern integrability theory (\cite{Liouville}, \cite{Anderson}, \cite{Ferapontov}, \cite{Dynnikov}, \cite{Adler}, \cite{Zhiber}, \cite{Ganzha}). Below we recall the main idea of the cascade method.

Consider a linear second order hyperbolic type PDE
\begin{equation}\label{I.1}
u_{xy}+a(x,y)u_x+b(x,y)u_y+c(x,y)u=0.
\end{equation}
Introduce new variables $u_{[1]}$ and $u_{[-1]}$ due to the relations (see, for example, \cite{Zhiber})
\begin{equation}\label{I.4}
\eqalign{
\left(\frac{\partial}{\partial y}+a\right)u=u_{[1]}, \qquad \left(\frac{\partial}{\partial x}+b\right)u_{[1]}=h_{[0]}u,\cr
\left(\frac{\partial}{\partial x}+b\right)u=u_{[-1]}, \qquad \left(\frac{\partial}{\partial y}+a\right)u_{[-1]}=k_{[0]}u.}
\end{equation}
Here functions $h_{[0]}$, $k_{[0]}$ defined due to the rule
\begin{equation}\label{I.2}
h_{[0]}=a_x+ab-c, \qquad k_{[0]}=b_y+ab-c
\end{equation}
are called Laplace invariants.
If the Laplace invariants do not vanish then the newly introduced functions solve linear PDE's as well
\begin{equation}\label{I.5}
\eqalign{
u_{[1]xy}+a_{[1]}u_{[1]x}+b_{[1]}u_{[1]y}+c_{[1]}u_{[1]}=0, \cr
u_{[-1]xy}+a_{[-1]}u_{[-1]x}+b_{[-1]}u_{[-1]y}+c_{[-1]}u_{[-1]}=0 }
\end{equation}
with the coefficients having the form
\begin{equation*}
a_{[1]}=a-(\ln h_{[0]})_y, \qquad b_{[1]}=b, \qquad c_{[1]}=a_{[1]}b_{[1]}+b_y-h_{[0]}
\end{equation*}
and
\begin{equation*}
a_{[-1]}=a, \qquad b_{[-1]}=b-(\ln k_{[0]})_x, \qquad c_{[-1]}=a_{[-1]}b_{[-1]}+a_x-k_{[0]}.
\end{equation*}
Transformations converting equation (\ref{I.1}) into equations (\ref{I.5}) are called Laplace y- and respectively Laplace x-transformations.

Repeated applications of these transformations create a sequence of the hyperbolic type equations referred to as  the Laplace sequence:
\begin{equation}\label{I.9}
\frac{\partial^2}{\partial x\partial y}u_{[i]}+a_{[i]}\frac{\partial}{\partial x}u_{[i]}+b_{[i]}\frac{\partial}{\partial y}u_{[i]}+c_{[i]}u_{[i]}=0, \quad \mbox{for}\quad i=0,1,2,\ldots
\end{equation}
and
\begin{equation}\label{I.10}
\fl \qquad \frac{\partial^2}{\partial x\partial y}u_{[-i]}+a_{[-i]}\frac{\partial}{\partial x}u_{[-i]}+b_{[-i]}\frac{\partial}{\partial y}u_{[-i]}+c_{[-i]}u_{[-i]}=0,\quad \mbox{for}\quad i=1,2,3,\ldots.
\end{equation}
It is assumed that $a_{[0]}=a$, $b_{[0]}=b$, $c_{[0]}=c$, $u_{[0]}=u$. The coefficients are given by
\begin{eqnarray}\label{I.11}
&a_{[i]}=a_{[i-1]}-(\ln h_{[i-1]})_y,\qquad & b_{[i]}=b_{[i-1]},\nonumber \\
&k_{[i]}=h_{[i-1]},  &c_{[i]}=a_{[i]}b_{[i]}+(b_{[i]})_y-h_{[i-1]},\\
&h_{[i]}=2h_{[i-1]}-k_{[i-1]}-(\ln h_{[i-1]})_{xy},\quad & \mbox{for}\quad i=0,1,2,\ldots \nonumber 
\end{eqnarray}
and by
\begin{eqnarray}\label{I.12}
\fl\quad &a_{[-i-1]}=a_{[-i]},  & b_{[-i-1]}=b_{[-i]}-(\ln k_{[-i]})_x, \nonumber \\
\fl\quad &h_{[-i-1]}=k_{[-i]}, & c_{[-i-1]}=a_{[-i-1]}b_{[-i-1]}+(a_{[-i]})_x-k_{[-i]},\\
\fl\quad &k_{[-i-1]}=2k_{[-i]}-h_{[-i]}-(\ln k_{[-i]})_{xy}, \quad & \mbox{for}\quad i=1,2,3,\ldots. \nonumber
\end{eqnarray}
The sequence of the functions $\left\{h_{[i]}\right\}_{i=-\infty}^{i=\infty}$ is called the sequence of the Laplace invariants.
Equations (\ref{I.9}) and (\ref{I.10}) can be represented as systems of the form
\begin{equation}\label{I.13}
\left(\frac{\partial}{\partial y}+a_{[i]}\right)u_{[i]}=u_{[i+1]}, \qquad \left(\frac{\partial}{\partial x}+b\right)u_{[i+1]}-h_{[i]}u_{[i]}=0
\end{equation}
and correspondingly
\begin{equation}\label{I.14}
\left(\frac{\partial}{\partial x}+b_{[-i]}\right)u_{[-i]}=u_{[-i-1]}, \qquad \left(\frac{\partial}{\partial y}+a\right)u_{[-i-1]}-k_{[-i]}u_{[-i]}=0.
\end{equation}

As it is obvious from the formulas above if $h_{[i_0]}=0$ for $i_0\geqslant 0$, then equation (\ref{I.9}) is not defined for 
$i\geqslant i_0+1$. Therefore in this case the Laplace sequence terminates on the right side. Similarly, if $k_{[-i_1]}=0$ for $i_1\geqslant 0$ then equation (\ref{I.10}) isn't defined for $i\geqslant i_1$, i.e. the sequence terminates on the left side.

A remarkable application of the construction is based on the following statement. If the sequence of the Laplace transformations for equation (\ref{I.1}) terminates on both sides then the equation is explicitly solved without any 
quadrature. Even in the case when the sequence terminates only on one side the Laplace cascade method allows one finding of a special solution depending on an arbitrary function.

In relatively recent studies (see \cite{Liouville}, \cite{Anderson}, \cite{Adler}, \cite{Zhiber}, \cite{Ganzha}) Laplace cascade method has successfully been extended to the case of nonlinear partial differential equations of Liouville type.

In the present article we discuss the question of what are the characteristic features of the Laplace transformation sequence
for the integrable nonlinear PDE of the  sine-Gordon type. Actually we concentrate on the equations of the form
\begin{equation}\label{Iy.1}
u_{xy}=F(u,u_x,u_y),
\end{equation}
which are integrated by means of the inverse scattering transform method (see, for instance, \cite{Novikovbook}), i.e. they admit hierarchies of generalized symmetries on each of the characteristic directions and has no nontrivial integrals.

Recall that the linear equation
\begin{equation}\label{Iy.2}
v_{xy}+av_x+bv_y+cv=0
\end{equation}
is called the linearization of the equation (\ref{Iy.1}) or, synonymously, its Fr\'{e}chet derivative if the coefficients of the equation are given by the formulas 
\begin{equation}\label{abc}
a=-\frac{\partial F}{\partial u_x}, \quad b=-\frac{\partial F}{\partial u_y}, \quad c=-\frac{\partial F}{\partial u}.
\end{equation}
We stress that now the coefficients of  (\ref{Iy.2}) depend on an arbitrary solution $u=u(x,y)$ of the equation (\ref{Iy.1}), that is considered as a functional parameter. Therefore we have to adapt formulas (\ref{I.2})-(\ref{I.14}) by replacing the operators of partial derivatives $\frac{\partial}{\partial x}$ and $\frac{\partial}{\partial y}$ by the operators $D_x$ and $D_y$ of the total derivative with respect to $x$ and $y$. For example the Laplace invariants for the linearized equation $h$ and $k$ are found due to the formulas
\begin{equation}\label{h0k0}
h=D_x(a)+ab-c, \qquad k=D_y(b)+ab-c.
\end{equation}
Assuming that $h_{[0]}:=h$, $k_{[0]}:=k$, $a_{[0]}:=a$, $b_{[0]}:=b$, $c_{[0]}:=c$, where function $a$, $b$ and $c$ are given by (\ref{abc}) we can rewrite all the necessary formulas (\ref{I.2})-(\ref{I.14}). Iterated application of the Laplace transformation to the linearized equation (\ref{Iy.2}) defines an infinite sequence of the equations for unknown functions $v_{[i]}$:
\begin{equation}\label{I.13-v}
\left(\frac{\partial}{\partial y}+a_{[i]}\right)v_{[i]}=v_{[i+1]}, \qquad \left(\frac{\partial}{\partial x}+b_{[0]}\right)v_{[i+1]}-h_{[i]}v_{[i]}=0
\end{equation}
and
\begin{equation}\label{I.14-v}
\fl \qquad \left(\frac{\partial}{\partial x}+b_{[-i]}\right)v_{[-i]}=v_{[-i-1]}, \qquad \left(\frac{\partial}{\partial y}+a_{[0]}\right)v_{[-i-1]}-k_{[-i]}v_{[-i]}=0,
\end{equation}
where the coefficients are determined due to the relations:
\begin{eqnarray}\label{ah+}
\eqalign{
a_{[1]}=a_{[0]}-D_y(\ln h_{[0]}), \quad & h_{[1]}=h_{[0]}+D_x(a_{[1]})-D_y(b_{[0]}),\\
a_{[2]}=a_{[1]}-D_y(\ln h_{[1]}), \quad & h_{[2]}=h_{[1]}+D_x(a_{[2]})-D_y(b_{[0]}),\\
\ldots\ldots\ldots\ldots,\\
a_{[i+1]}=a_{[i]}-D_y(\ln h_{[i]}), \quad & h_{[i+1]}=h_{[i]}+D_x(a_{[i+1]})-D_y(b_{[0]}),\\
\ldots\ldots\ldots\ldots.}
\end{eqnarray}
By using $h_{[-1]}=k_{[0]}$ we can find coefficients with negative indices:
\begin{eqnarray}\label{bh-}
\fl \qquad \eqalign{
b_{[-1]}=b_{[0]}-D_x(\ln h_{[-1]}), \quad & h_{[-2]}=h_{[-1]}-D_x(a_{[0]})+D_y(b_{[-1]}),\\
b_{[-2]}=b_{[-1]}-D_x(\ln h_{[-2]}), \quad & h_{[-3]}=h_{[-2]}-D_x(a_{[0]})+D_y(b_{[-2]}),\\
\ldots\ldots\ldots\ldots,\\
b_{[-i-1]}=b_{[-i]}-D_x(\ln h_{[-i-1]}), \quad & h_{[-i-2]}=h_{[-i-1]}-D_x(a_{[0]})+D_y(b_{[-i-1]}),\\
\ldots\ldots\ldots\ldots.}
\end{eqnarray}

In this article we continue the study begun in our papers \cite{HabKhaPo}, \cite{HabKhaJPA17}, \cite{HabKhaJMS21}. We conjecture that for any integrable equation of the form (\ref{Iy.1}) that has no nontrivial characteristic integrals the sequence of the Laplace transformations associated to its linearization admits a finite field reduction, i.e. it is compatible with a properly chosen constraint of the form
\begin{equation}\label{Iy.12}
\eqalign{
v_{[m+1]}=\alpha_{[-k]}v_{[-k]}+\alpha_{[-k+1]}v_{[-k+1]}+\cdots+\alpha_{[m]}v_{[m]},\cr
v_{[-k-1]}=\beta_{[-k]}v_{[-k]}+\beta_{[-k+1]}v_{[-k+1]}+\cdots+\beta_{[m]}v_{[m]}.}
\end{equation}
Where indices are some integers, such that $m\geqslant1$, $k\geqslant1$; coefficients $\alpha_j, \beta_j$ depend on the dynamical variables $u$, $u_x$, $u_y, \ldots$ and on a spectral parameter $\lambda$. In addition it is required, that $\alpha_{[-k]}$ and $\beta_{[m]}$ don't vanish identically.

{\bf Proposition 1.} Assume that the Laplace invariants $h_{[i]}$ don't vanish for any integer $i$ and the Laplace sequence (\ref{I.13-v}), (\ref{I.14-v}) is compatible with the constraint (\ref{Iy.12}). Then for arbitrary integer $j$ function $v_{[j]}$ can be represented as follows
\begin{equation*}
v_{[j]}=\alpha^{j}_{[-k]}v_{[-k]}+\alpha^{j}_{[-k+1]}v_{[-k+1]}+\cdots+\alpha^{j}_{[m]}v_{[m]}.
\end{equation*}
In other words the constraint (\ref{Iy.12}) defines a reduction of the whole Laplace sequence.

{\bf Sketch of the proof.} By applying the operator $\left(D_y+a_{[m+1]}\right)$ to the first equation in (\ref{Iy.12}) we get the representation
\begin{equation*}
v_{[m+2]}=\beta^{1}_{[-k]}v_{[-k]}+\beta^{1}_{[-k+1]}v_{[-k+1]}+\cdots+\beta^{1}_{[m+1]}v_{[m+1]}.
\end{equation*}
Now the variable $v_{[m+1]}$ is easily excluded by virtue of (\ref{Iy.12}). Finally we obtain
\begin{equation*}
v_{[m+2]}=\alpha^{1}_{[-k]}v_{[-k]}+\alpha^{1}_{[-k+1]}v_{[-k+1]}+\cdots+\alpha^{1}_{[m]}v_{[m]}.
\end{equation*}
Continuing this way we find the desired representation for any $v_{[m+s]}$ with $s\geq 2$. In a similar way the case of the negative direction is studied. For example to get the constraint for $v_{[-k-2]}$ we apply to the second part of (\ref{Iy.12}) operator $\left(D_x+b_{[-k-1]}\right)$.

{\bf Example.} Below we discuss the Laplace sequence for the sine-Gordon equation
\begin{equation}\label{Iy.3}
u_{xy}=\sin u
\end{equation}
for which in \cite{HabKhaPo} we have found a finite field reduction and illustrated some of its applications.
Obviously linearization of (\ref{Iy.3}) looks like
\begin{equation}\label{Iy.4}
v_{xy}-\cos uv=0.
\end{equation}
According to the formulas (\ref{h0k0}), (\ref{ah+}), (\ref{bh-}) we have $h_{[0]}=h_{[-1]}=k_{[0]}=k_{[1]}=\cos u$, $h_{[1]}=h_{[-2]}=k_{[2]}=k_{[-1]}=\frac{1}{\cos u}+\frac{u_xu_y}{\cos^2 u}$.
Let us give in an explicit form several first members of the Laplace sequence
\begin{equation}\label{Iy.5}
\eqalign{
v_{[2]xy}+2u_y\tan uv_{[2]x}-\left(\frac{1}{\cos u}+\frac{u_xu_y}{\cos^2 u}\right)v_{[2]}=0,\cr
v_{[1]xy}+u_y\tan uv_{[1]x}-\cos uv_{[1]}=0,\cr
v_{[0]xy}-\cos uv_{[0]}=0,\cr
v_{[-1]xy}+u_x\tan uv_{[-1]y}-\cos uv_{[-1]}=0,\cr
v_{[-2]xy}+2u_x\tan uv_{[-2]y}-\left(\frac{1}{\cos u}+\frac{u_xu_y}{\cos^2 u}\right)v_{[-2]}=0.}
\end{equation}
The following statement has been proved in \cite{HabKhaPo}.

{\bf Theorem 1.}
System of the hyperbolic equations (\ref{Iy.5}) is compatible with the pair of constraints
\begin{equation}\label{Iy.10}
\eqalign{
v_{[2]}=\alpha_{[-1]}v_{[-1]}+\alpha_{[0]}v_{[0]}+\alpha_{[1]}v_{[1]},  \cr
v_{[-2]}=\beta_{[-1]}v_{[-1]}+\beta_{[0]}v_{[0]}+\beta_{[1]}v_{[1]},}
\end{equation}
where the coefficients are given by
\begin{eqnarray*}
\alpha_{[-1]}&=-\lambda\frac{u_y}{\sin u}, \qquad \alpha_{[0]}=\lambda, \qquad \alpha_{[1]}=\frac{u_y}{\cos u\sin u},\\
\beta_{[-1]}&=\frac{u_x}{\cos u\sin u}, \qquad \beta_{[0]}=\frac{1}{\lambda}, \qquad  \beta_{[1]}=-\frac{u_x}{\lambda\sin u},
\end{eqnarray*}
$\lambda$ is an arbitrary constant.

A remarkable corollary of the Theorem 1 claims that system of the equations (\ref{Iy.10}) generates an unusual Lax pair for the sine-Gordon equation realized in $3\times 3$ matrices. By a very elementary transformation we deduce from this Lax pair the recursion operators, corresponding to the hierarchies of symmetries in the directions of $x$ and $y$ (see \cite{HabKhaPo}). By a direct computation one reduces the order of this Lax pair and arrive at a system of three nonlinear equations from which Dubrovin type equation can be derived (\cite{HabKhaJPA20}, \cite{UMJ21}, \cite{Non23}), suitable for constructing algebro-geometric solutions (about algebro-geometric solutions see \cite{DubrovinMatveevNovikov}). 

By passing to new variables in a proper way one can derive from the triple of nonlinear equations the standard Lax pair (see \cite{HabKhaPo}, \cite{HabKhaJPA17}, \cite{HabKhaJMS21}). 

In our previous studies (\cite{HabKhaPo}, \cite{HabKhaJPA17}, \cite{HabKhaJMS21}), we have already tested the hypothesis in addition to the sine-Gordon equation also using the example of integrable equation $u_{xy}=f(u)\sqrt{1+u^2_x}$, where $f$ is a solution of the equation $f''=\gamma f$, $\gamma=const$, found in~\cite{SokMesh}.

In the present paper the following two equations presented in the list of \cite{SokMesh}:
\begin{equation}\label{1.1}
u_{xy}=\sqrt{u_x}\sqrt{u_y^2+1}
\end{equation}
and
\begin{equation}\label{2.1}
u_{xy}=\sqrt{u_x^2+1}\sqrt{u_y^2+1}
\end{equation}
are investigated.

Let us briefly discuss the content of the article. In \S2 we apply the scheme proposed in Introduction to equations (\ref{1.1}) and (\ref{2.1}). For linearizations of these equations we have found finite reductions of the form (\ref{Iy.12}) with $k=1$ and $m=1$ (see Theorems 2 and 3). It is shown that the finite reductions can be rewritten as $3\times 3$ Lax pairs for the original nonlinear equations. By using these unusual Lax pairs we derived the recursion operators for both equations describing hierarchies of higher symmetries corresponding to characteristic directions $x$ and $y$. In section 3 the order for the $3\times 3$ Lax pairs is reduced by one and systems of nonlinear equations are obtained which are important for constructing algebro-geometric solutions to the equations. After appropriate change of the variables the nonlinear systems convert to $2\times 2$ Lax pairs of the usual type. To the best of our knowledge recursion operators and Lax pairs for (\ref{1.1}), (\ref{2.1}) have not been found before.

\section{Reductions of the Laplace sequence for the sine-Gordon type models}

In this section we determine the Laplace sequence corresponding to equations (\ref{1.1}) and (\ref{2.1}) and then find finite reductions connected with the integrability property of the equations.

\subsection{Finding finite reductions for equation (\ref{1.1})}

Let us first concentrate on equation (\ref{1.1}) for which the coefficients of the linearized equation (\ref{Iy.2}) are 
\begin{equation}\label{c1}
a=-\frac{\sqrt{u_y^2+1}}{2\sqrt{u_x}}, \quad b=-\frac{u_y\sqrt{u_x}}{\sqrt{u_y^2+1}}, \quad c=0.
\end{equation}
Let us consider a finite subsystem
\begin{equation}\label{subsys-1}
\eqalign{
\left(\frac{\partial}{\partial y}+a_{[1]}\right)v_{[1]}=v_{[2]}, \qquad \quad &\left(\frac{\partial}{\partial x}+b_{[0]}\right)v_{[2]}-h_{[1]}v_{[1]}=0,\cr
\left(\frac{\partial}{\partial y}+a_{[0]}\right)v_{[0]}=v_{[1]}, \qquad &\left(\frac{\partial}{\partial x}+b_{[0]}\right)v_{[1]}-h_{[0]}v_{[0]}=0,\cr
\left(\frac{\partial}{\partial x}+b_{[0]}\right)v_{[0]}=v_{[-1]}, \qquad &\left(\frac{\partial}{\partial y}+a_{[0]}\right)v_{[-1]}-k_{[0]}v_{[0]}=0,\cr
\left(\frac{\partial}{\partial x}+b_{[-1]}\right)v_{[-1]}=v_{[-2]}, \qquad &\left(\frac{\partial}{\partial y}+a_{[0]}\right)v_{[-2]}-k_{[-1]}v_{[-1]}=0
}
\end{equation}
of the infinite system of equations (\ref{I.13-v}), (\ref{I.14-v}) generated by the Laplace transformations applied to the linearization of the equation (\ref{1.1}). Here the coefficients $a_{[0]}$, $b_{[0]}$ are given by (\ref{c1}), and the coefficients $h_{[0]}$, $k_{[0]}$, $a_{[1]}$, $b_{[-1]}$, $h_{[1]}$, $k_{[-1]}$ are determined by the formulas (\ref{h0k0}), (\ref{ah+}) and (\ref{bh-}).

{\bf Theorem 2.} System of the linear equations (\ref{subsys-1}) corresponding to equation (\ref{1.1}) is consistent with the following constraint
\begin{equation}\label{1.9}
\eqalign{
v_{[2]}=\alpha_{[-1]}v_{[-1]}+\alpha_{[0]}v_{[0]}+\alpha_{[1]}v_{[1]},\cr
v_{[-2]}=\beta_{[-1]}v_{[-1]}+\beta_{[0]}v_{[0]}+\beta_{[1]}v_{[1]} }
\end{equation}
if the coefficients are defined as follows 
\begin{equation}\label{1.10}
\eqalign{
\alpha_{[-1]}=\frac{\lambda u_y\sqrt{u_y^2+1}}{\sqrt{u_x}}, \quad \alpha_{[0]}=-\lambda, \qquad \alpha_{[1]}=-\frac{u_xu_y}{u_{xx}}, \cr
\beta_{[-1]}=-\frac{u_y^2+1}{2u_{yy}}, \qquad \quad \beta_{[0]}=\frac{1}{4\lambda}, \qquad \beta_{[1]}=\frac{\sqrt{u_x}}{2\lambda\sqrt{u_y^2+1}},}
\end{equation}
where $\lambda$ is a complex parameter.

{\bf Proof.} Substituting representations (\ref{1.9}) into equations (\ref{subsys-1}) we obtain the following six equations for determining unknown coefficients $\alpha_{[i]}$, $\beta_{[i]}$:
\begin{equation}\label{1.11}
\eqalign{
v_{[1]y}=v_{[2]}-a_{[1]}v_{[1]}=(\alpha_{[1]}-a_{[1]})v_{[1]}+\alpha_{[0]}v_{[0]}+\alpha_{[-1]}v_{[-1]},\cr
v_{[0]y}=v_{[1]}-a_{[0]}v_{[0]},\cr
v_{[-1]y}=h_{[-1]}v_{[0]}-a_{[0]}v_{[-1]}}
\end{equation}
and
\begin{equation}\label{1.12}
\eqalign{
v_{[1]x}=h_{[0]}v_{[0]}-b_{[0]}v_{[1]},\cr
v_{[0]x}=v_{[-1]}-b_{[0]}v_{[0]},\cr
v_{[-1]x}=v_{[-2]}-b_{[-1]}v_{[-1]}=(\beta_{[-1]}-b_{[-1]})v_{[-1]}+\beta_{[0]}v_{[0]}+\beta_{[1]}v_{[1]}.}
\end{equation}
We do not explicitly present the remaining two equations because they do not add conditions for determining the required coefficients.

The compatibility conditions $(v_{[i]x})_y=(v_{[i]y})_x$ imply a system of nonlinear equations for sought functions 
\begin{equation}\label{1.13}
\eqalign{
D_y(\beta_{[-1]})+\beta_{[1]}\alpha_{[-1]}-h_{[-2]}=0,\cr
D_y(\beta_{[0]})+\beta_{[1]}\alpha_{[0]}+\beta_{[-1]}h_{[-1]}=0,\cr
D_y(\beta_{[1]})+\beta_{[1]}(\alpha_{[1]}+a_{[0]}-a_{[1]})+\beta_{[0]}=0,\cr
D_x(\alpha_{[-1]})+\alpha_{[-1]}(\beta_{[-1]}+b_{[0]}-b_{[-1]})+\alpha_{[0]}=0,\cr
D_x(\alpha_{[0]})+\alpha_{[-1]}\beta_{[0]}+\alpha_{[1]}h_{[0]}=0,\cr
D_x(\alpha_{[1]})+\alpha_{[-1]}\beta_{[1]}-h_{[1]}=0.}
\end{equation}
Here the coefficients are given in an explicit form 
\begin{equation}\label{1.14}
\eqalign{
h_{[0]}=\frac{\sqrt{u_y^2+1}u_{xx}}{4u_x^{3/2}}, &h_{[1]}=\frac{u_xu_yu_{xxx}}{u_{xx}^2}-\frac{u_y}{2}-\frac{u_x^{3/2}\sqrt{u_y^2+1}}{u_{xx}},\cr
h_{[-1]}=-\frac{\sqrt{u_x}u_{yy}}{(u_y^2+1)^{3/2}}, \qquad &h_{[-2]}=\frac{(u_y^2+1)u_{yyy}}{2u_{yy}^2}-\frac{u_y}{2},\cr
a_{[0]}=-\frac{\sqrt{u_y^2+1}}{2\sqrt{u_x}}, &a_{[1]}=-\frac{u_yu_{yy}}{u_y^2+1}+\frac{\sqrt{u_y^2+1}}{2\sqrt{u_x}}-\frac{u_xu_y}{u_{xx}},\cr
b_{[0]}=-\frac{u_y\sqrt{u_x}}{\sqrt{u_y^2+1}}, &b_{[-1]}=\frac{u_y\sqrt{u_x}}{\sqrt{u_y^2+1}}-\frac{u_{xx}}{2u_x}
-\frac{u_y^2+1}{2u_{yy}}.}
\end{equation}
The preliminary analysis proves that functions $\alpha_{[i]}$, $\beta_{[i]}$ for $i=-1,0,1$ depend on dynamical variables $u_x$, $u_y$, $u_{xx}$, $u_{yy}$ as follows
\begin{equation*}
\eqalign{
\alpha_{[1]}=\alpha_{[1]}(u_x, u_y, u_{xx}), \qquad
&\beta_{[1]}=\beta_{[1]}(u_x, u_y),\cr
\alpha_{[0]}=\alpha_{[0]}(u_x, u_y), 
\qquad &\beta_{[0]}=\beta_{[0]}(u_x, u_y),\cr
\alpha_{[-1]}=\alpha_{[-1]}(u_x, u_y), 
&\beta_{[-1]}=\beta_{[-1]}(u_x, u_y,u_{yy}).}
\end{equation*}
Let's substitute these representations of the desired functions into the system of equations (\ref{1.13}).
Further, collecting the coefficients of the independent variables $u_{xxx}$ and $u_{yyy}$ in the first and last equations of system (\ref{1.13}), we obtain equations for determining $\alpha_{[1]}$, $\beta_{[-1]}$
\begin{equation*}
\eqalign{
\alpha_{[1]u_{xx}}-\frac{u_xu_y}{u_{xx}^2}=0,\cr
\beta_{[-1]u_{yy}}-\frac{u_y^2+1}{2u_{yy}^2}=0,}
\end{equation*}
which immediately give rise to
\begin{equation*}
\eqalign{
\alpha_{[1]}=-\frac{u_xu_y}{u_{xx}}+\tilde\alpha_{[1]}(u_x,u_y),\cr
\beta_{[-1]}=-\frac{u_y^2+1}{2u_{yy}}+\tilde\beta_{[-1]}(u_x,u_y),}
\end{equation*}
where $\tilde\alpha_{[1]}$ and $\tilde\beta_{[-1]}$ are functions to be found. By taking into account the expressions found above one can rewrite the first and last equations in (\ref{1.13}) in the following form
\begin{equation}\label{1.15}
\eqalign{
\tilde\beta_{[-1]u_y}u_{yy}+\tilde\beta_{[-1]u_x}\sqrt{u_x}\sqrt{u_y^2+1}-\frac{u_y}{2}+\beta_{[1]}\alpha_{[-1]}=0,\cr
\tilde\alpha_{[1]u_x}u_{xx}+\tilde\alpha_{[1]u_y}\sqrt{u_x}\sqrt{u_y^2+1}-\frac{u_y}{2}+\beta_{[1]}\alpha_{[-1]}=0.}
\end{equation}
In the latter equations the variables $u_{xx}$ and $u_{yy}$ might be considered as independent ones. Therefore, we can conclude that each of the following functions depends on only one variable $\tilde\beta_{[-1]}=\tilde\beta_{[-1]}(u_x)$ and $\tilde\alpha_{[ 1]}=\tilde\alpha_{ [1]}(u_y)$.
Further, we subtract one of the equations in (\ref{1.15}) from another and get
\begin{equation*}
\tilde\beta_{[-1]u_x}=\tilde\alpha_{[1]u_y}
\end{equation*}
that gives $\tilde\beta_{[-1]u_x}=\tilde\alpha_{[1]u_y}=C_1=const$. Thus we have
\begin{equation*}
\eqalign{
\tilde\beta_{[-1]}=C_1u_x+C_2, \qquad
&\tilde\alpha_{[1]}=C_1u_y+C_3,}
\end{equation*}
where $C_2$ and $C_3$ are some constants.
Consequently functions $\alpha_{[1]}$ and $\beta_{[-1]}$ are specified up to constant parameters:
\begin{equation*}
\eqalign{
\alpha_{[1]}=-\frac{u_xu_y}{u_{xx}}+C_1u_y+C_3,\cr
\beta_{[-1]}=-\frac{u_y^2+1}{2u_{yy}}+C_1u_x+C_2.}
\end{equation*}

Let us concentrate on the remaining four equations of system (\ref{1.13}). Comparing in these equations the coefficients of the independent variables $u_{xx}$ and $u_{yy}$ we obtain equations
\begin{equation*}
\eqalign{
\beta_{[0]u_y}(u_y^2+1)^{3/2}-\sqrt{u_x}(C_1u_x+C_2)=0,\cr
\beta_{[1]u_y}(u_y^2+1)+\beta_{[1]}u_y=0,\cr
2\alpha_{[-1]u_x}u_x+\alpha_{[-1]}=0,\cr
4\alpha_{[0]u_x}u_x^{3/2}+\sqrt{u_y^2+1}(C_1u_y+C_3)=0.}
\end{equation*}
It is obvious that these equations are effectively solved and give the expressions
\begin{equation*}
\eqalign{
\beta_{[0]}=\frac{(C_1u_x+C_2)\sqrt{u_x}u_y}{\sqrt{u_y^2+1}}+\tilde{\beta}_{[0]}(u_x),\cr
\beta_{[1]}=\frac{\tilde{\beta}_{[1]}(u_x)}{\sqrt{u_y^2+1}},\cr
\alpha_{[-1]}=\frac{\tilde{\alpha}_{[-1]}(u_y)}{\sqrt{u_x}},\cr
\alpha_{[0]}=\frac{(C_1u_y+C_3)\sqrt{u_y^2+1}}{2\sqrt{u_x}}+\tilde{\alpha}_{[0]}(u_y),}
\end{equation*}
where $\tilde{\alpha}_{[i]}(u_y)$ and $\tilde{\beta}_{[j]}(u_x)$ are some functions to be determined.

Let us specify the third and fourth equations of system (\ref{1.13}) by virtue of found expressions of the functions $\alpha_{[1]}$ and $\beta_{[-1]}$:
\begin{equation}\label{1.16}
\eqalign{
\frac{\sqrt{u_y^2+1}}{\sqrt{u_x}}&\left(\tilde\beta_{[1]u_x}u_x
+\tilde\beta_{[0]}\sqrt{u_x}-\tilde\beta_{[1]}\right)+\cr
&+\left(C_1\tilde\beta_{[1]}+C_1u_x^{3/2}+C_2\sqrt{u_x}\right)u_y+C_3\tilde\beta_{[1]}=0,}
\end{equation}
\begin{equation}\label{1.17}
\eqalign{
\frac{\sqrt{u_x}}{\sqrt{u_y^2+1}}&\left(\tilde\alpha_{[-1]u_y}(u_y^2+1)
-2\tilde\alpha_{[-1]}u_y+\tilde\alpha_{[0]}\sqrt{u_y^2+1}\right)+\cr
&+C_1\tilde\alpha_{[-1]}u_x+C_2\tilde\alpha_{[-1]}+\frac{\sqrt{u_y^2+1}}{2}(C_1u_y+C_3)=0.}
\end{equation}

Note that any of these two equations splits down into three equations since in equation (\ref{1.16}) the variables $u_y$ and $\sqrt{u_y^2+1}$ can be considered as independent ones, similarly in equation (\ref{1.17}) the variables $u_x$ and $\sqrt{u_x}$ are linearly independent. Let us write down all these six equations:
\begin{equation*}
\eqalign{
C_1\tilde\beta_{[1]}+C_1u_x^{3/2}+C_2\sqrt{u_x}=0,\cr
\tilde\beta_{[1]u_x}u_x+\tilde\beta_{[0]}\sqrt{u_x}-\tilde\beta_{[1]}=0,\cr
C_3\tilde\beta_{[1]}=0,\cr
\tilde\alpha_{[-1]u_y}(u_y^2+1)-2\tilde\alpha_{[-1]}u_y+\tilde\alpha_{[0]}\sqrt{u_y^2+1}=0,\cr
C_1\tilde\alpha_{[-1]}=0,\cr
C_2\tilde\alpha_{[-1]}+\frac{\sqrt{u_y^2+1}}{2}(C_1u_y+C_3)=0.}
\end{equation*}
It follows from the third and fifth equations that $C_1=0$, $C_3=0$ because $\tilde\beta_{[1]}$ and $\tilde\alpha_{[-1]}$ don't vanish, then the first equation implies $C_2=0$.

Now we turn back to the first equation in (\ref{1.13}). Due to the reasoning above it takes the form
\begin{equation}\label{1.18}
\frac{\tilde\beta_{[1]}\tilde\alpha_{[-1]}}{\sqrt{u_x}\sqrt{u_y^2+1}}-\frac{u_y}{2}=0.
\end{equation}
Evidently the equation admits a separation of the variables
\begin{equation*}
\frac{\tilde\alpha_{[-1]}(u_y)}{u_y\sqrt{u_y^2+1}}=\frac{\sqrt{u_x}}{2\tilde\beta_{[1]}(u_x)}=C_4=const.
\end{equation*}
Therefore the sought functions $\tilde\alpha_{[-1]}$ and $\tilde\beta_{[1]}$ are of the form
\begin{equation*}
\tilde\alpha_{[-1]}(u_y)=C_4u_y\sqrt{u_y^2+1},\qquad
\tilde\beta_{[1]}(u_x)=\frac{\sqrt{u_x}}{2C_4}.
\end{equation*}
After all of the specifications equations (\ref{1.16}) and (\ref{1.17}) are essentially simplified and get the form
\begin{equation*}
\tilde\beta_{[0]}(u_x)=\frac{1}{4C_4},\qquad
\tilde\alpha_{[0]}(u_y)=-C_4.
\end{equation*}
Thus all of the functions are found, they are of the form (\ref{1.10}), where $\lambda:=C_4$. Theorem~2 is proved.

Here we discuss some useful consequences of Theorem 2. Obviously systems of equations (\ref{1.11}), (\ref{1.12}) can be represented as a pair of systems 
\begin{equation}\label{1.28}
\Psi_x=A\Psi, \qquad \Psi_y=B\Psi
\end{equation}
for the vector $\Psi=(v_{[1]},v_{[0]},v_{[-1]})^T$, where
\begin{eqnarray}\label{1.29}
\eqalign{
A=\left(
\begin{array}{ccc}
\frac{u_y\sqrt{u_x}}{\sqrt{u_y^2+1}} & \frac{\sqrt{u_y^2+1}u_{xx}}{4u_x^{3/2}} & 0\\
0 & \frac{u_y\sqrt{u_x}}{\sqrt{u_y^2+1}} & 1\\
\frac{\sqrt{u_x}}{2\lambda\sqrt{u_y^2+1}} & \frac{1}{4\lambda} & -\frac{u_y^2+1}{2u_{yy}}+\frac{u_y\sqrt{u_x}}{\sqrt{u_y^2+1}}
\end{array}
\right), \cr
B=\left(
\begin{array}{ccc}
\frac{u_yu_{yy}}{u_y^2+1}-\frac{\sqrt{u_y^2+1}}{2\sqrt{u_x}} & -\lambda & \frac{\lambda u_y\sqrt{u_y^2+1}}{\sqrt{u_x}}\\
1 & \frac{\sqrt{u_y^2+1}}{2\sqrt{u_x}} & 0\\
0 & -\frac{\sqrt{u_x}u_{yy}}{(u_y^2+1)^{3/2}} & \frac{\sqrt{u_y^2+1}}{2\sqrt{u_x}} 
\end{array}
\right).}
\end{eqnarray}

{\bf Corollary of Theorem 2.} System of equations (\ref{1.28}) is compatible if and only if function $u=u(x,y)$ is a solution to the equation (\ref{1.1}). 

In other words a pair of the linear equations given in (\ref{1.28}) determines the Lax pair for equation (\ref{1.1}). This Lax pair has a very unusual form: it is realized in $3\times 3$ matrices and the spectral parameter $\lambda$ enters in a rather simple way. One of its peculiarities is connected with the recursion operator. To derive the recursion operator we represent the matrix equations in (\ref{1.28}) as the third order scalar differential equations for unknown~$v_{[0]}$:
\begin{equation}\label{3.5-oim}
 v_{[0]xxx}-\frac{u_{xx}}{u_x}v_{[0]xx}-\left(u_x+\frac{u_{xxx}}{2u_x}-\frac{3u_{xx}^2}{4u_x^2}+\frac{1}{4\lambda}\right)v_{[0]x}-\frac{u_{xx}}{2}v_{[0]}=0
\end{equation} 
and, respectively
\begin{equation}\label{3.1-oim}
\fl v_{[0]yyy}-\left(\frac{u_{yy}}{u_y}+\frac{2u_yu_{yy}}{u_y^2+1}\right)v_{[0]yy}-\left(\frac{u_yu_{yyy}}{u_y^2+1}-
\frac{3u_y^2u_{yy}^2}{(u_y^2+1)^2}-\lambda\right)v_{[0]y}-\lambda \frac{u_{yy}}{u_y}v_{[0]}=0.
\end{equation}
Let us rewrite equations (\ref{3.5-oim}) and (\ref{3.1-oim}) in the following operator form: 
\begin{equation}\label{3.5}
\fl \left(D_x^3-\frac{u_{xx}}{u_x}D_x^2-\left(u_x+\frac{u_{xxx}}{2u_x}-\frac{3u_{xx}^2}{4u_x^2}\right)D_x-\frac{u_{xx}}{2}\right)v_{[0]}=\frac{1}{4\lambda} D_x\left(v_{[0]}\right),
\end{equation}
\begin{equation}\label{3.1}
\fl \left(D_y^3-\frac{(3u_y^2+1)u_{yy}}{u_y(u_y^2+1)}D_y^2-\left(\frac{u_yu_{yyy}}{u_y^2+1}-
\frac{3u_y^2u_{yy}^2}{(u_y^2+1)^2}\right)D_y\right)v_{[0]}=-\lambda u_yD_y\left(\frac{1}{u_y}v_{[0]}\right).
\end{equation}
Afterwards we multiply (\ref{3.5}) by $D_x^{-1}$ and, respectively, multiply (\ref{3.1}) by $u_yD_y^{-1}\frac{1}{u_y}$ and get the following relations 
\begin{equation}\label{3.2}
R_xv_{[0]}=\frac{1}{4\lambda} v_{[0]} \quad \mbox{and} \quad R_yv_{[0]}=-\lambda v_{[0]},
\end{equation}
where $R_x$ and $R_y$ are the recursion operators:
\begin{eqnarray}\label{R-1}
\eqalign{
&\fl \qquad R_x=D_x^2-\frac{u_{xx}}{u_x}D_x-\left(u_x+\frac{u_{xx}^2}{4u_x^2}\right)+D_x^{-1}\left(\frac{u_{xx}}{2}+\frac{u_{xx}u_{xxx}}{u_x^2}-\frac{u_{xx}^3}{2u_x^3}\right),\\
&\fl R_y=D_y^2-\frac{2u_yu_{yy}}{u_y^2+1}D_y-\frac{u_y^2u_{yy}^2}{(u_y^2+1)^2}
+u_yD_y^{-1}\left(\frac{u_{yy}u_{yyy}(3u_y^2+1)}{u_y(u_y^2+1)^2}-\frac{(3u_y^2-1)u_{yy}^3}{(u_y^2+1)^3}\right).}
\end{eqnarray}

{\bf Proposition 2.} Operators $R_x$ and $R_y$ given in (\ref{R-1}) define recursion operators generating hierarchies of the symmetries in the directions of $x$ and respectively $y$.

For example by applying these operators to the generators of classical symmetries $u_x$ and $u_y$ we get 
\begin{eqnarray*}
& R_x(u_x)=u_{xxx}-\frac{3u^2_{xx}}{4u_x}-\frac{3}{4}u_x^2,\\
& R_y(u_y)=u_{yyy}-\frac{3u_yu^2_{yy}}{2(u^2_y+1)}.
\end{eqnarray*}
Thus the first members of the hierarchies are: 
$$u_t=u_{xxx}-\frac{3u^2_{xx}}{4u_x}-\frac{3}{4}u_x^2, \qquad u_\tau=u_{yyy}-\frac{3u_yu^2_{yy}}{2(u^2_y+1)}.$$

\subsection{Reductions for equation (\ref{2.1})}

Let us turn to equation (\ref{2.1}). At first we find its linearization 
\begin{equation}\label{2.2}
v_{xy}-\frac{u_x\sqrt{u_y^2+1}}{2\sqrt{u_x^2+1}}v_x-\frac{u_y\sqrt{u_x^2+1}}{\sqrt{u_y^2+1}}v_y=0.
\end{equation}
Then, according to the above scheme, we determine sequence of the hyperbolic type equations due to the formulas (\ref{subsys-1}).

{\bf Theorem 3.} System of the linear equations (\ref{subsys-1}), corresponding to nonlinear equation (\ref{2.1}), is consistent with the constraints of the form 
\begin{equation}\label{1.9-2}
\eqalign{
v_{[2]}=\alpha_{[-1]}v_{[-1]}+\alpha_{[0]}v_{[0]}+\alpha_{[1]}v_{[1]},\cr
v_{[-2]}=\beta_{[-1]}v_{[-1]}+\beta_{[0]}v_{[0]}+\beta_{[1]}v_{[1]} }
\end{equation}
with the coefficients defined as follows:
\begin{equation}\label{2.4}
\eqalign{
\alpha_{[-1]}=\frac{\lambda u_y\sqrt{u_y^2+1}}{\sqrt{u_x^2+1}}, \qquad \alpha_{[0]}=-\lambda, \qquad \alpha_{[1]}=-\frac{u_y(u_x^2+1)}{u_{xx}}, \cr
\beta_{[-1]}=-\frac{u_x(u_y^2+1)}{u_{yy}}, \quad \quad \beta_{[0]}=-\frac{1}{\lambda}, \qquad \beta_{[1]}=\frac{u_x\sqrt{u_x^2+1}}{\lambda\sqrt{u_y^2+1}}.}
\end{equation}

The proof of Theorem 3 is quite similar to the proof of Theorem 2, so we don't present it here. It can be shown that using Theorem 3 one can obtain a Lax pair of the form (\ref{1.28}) for (\ref{2.1}) implemented in $3\times 3$ matrices:
\begin{eqnarray}\label{2.17}
\eqalign{
&A=\left(
\begin{array}{ccc}
\frac{u_y\sqrt{u_x^2+1}}{\sqrt{u_y^2+1}} & -\frac{\sqrt{u_y^2+1}u_{xx}}{(u_x^2+1)^{3/2}} & 0\\
0 & \frac{u_y\sqrt{u_x^2+1}}{\sqrt{u_y^2+1}} & 1\\
\frac{u_x\sqrt{u_x^2+1}}{\lambda\sqrt{u_y^2+1}} & \frac{1}{\lambda} & -\frac{u_x(u_y^2+1)}{u_{yy}}+\frac{u_y\sqrt{u_x^2+1}}{\sqrt{u_y^2+1}}
\end{array}
\right), \cr
&B=\left(
\begin{array}{ccc}
\frac{u_yu_{yy}}{u_y^2+1}-\frac{u_x\sqrt{u_y^2+1}}{\sqrt{u_x^2+1}} & -\lambda & \frac{\lambda u_y\sqrt{u_y^2+1}}{\sqrt{u_x^2+1}}\\
1 & \frac{u_x\sqrt{u_y^2+1}}{\sqrt{u_x^2+1}} & 0\\
0 & -\frac{\sqrt{u_x^2+1}u_{yy}}{(u_y^2+1)^{3/2}} & \frac{u_x\sqrt{u_y^2+1}}{\sqrt{u_x^2+1}} 
\end{array}
\right).}
\end{eqnarray}
One can immediately find the recursion operators for equation (\ref{2.1}) from the Lax pair (\ref{1.28}), (\ref{2.17}) by passing from the systems to the third order scalar differential equations for unknown $v_{[0]}$:
\begin{equation}\label{oim2-x}
\eqalign{
v_{[0]xxx}-\frac{(3u_x^2+1)u_{xx}}{u_x(u_x^2+1)}v_{[0]xx}-\cr
\quad -\left(u_x^2+1+\frac{u_xu_{xxx}}{u_x^2+1}-
\frac{3u_x^2u_{xx}^2}{(u_x^2+1)^2}-\lambda\right)v_{[0]x}-(\lambda-1) \frac{u_{xx}}{u_x}v_{[0]}=0}
\end{equation}
and
\begin{equation}\label{oim2-y}
\eqalign{
v_{[0]yyy}-\frac{(3u_y^2+1)u_{yy}}{u_y(u_y^2+1)}v_{[0]yy}-\cr
\quad 
-\left(u_y^2+1+\frac{u_yu_{yyy}}{u_y^2+1}-
\frac{3u_y^2u_{yy}^2}{(u_y^2+1)^2}-\lambda\right)v_{[0]y}-(\lambda-1) \frac{u_{yy}}{u_y}v_{[0]}=0}
\end{equation}
from which we derive the recursion operators $R_x$, $R_y$:
\begin{eqnarray}\label{R-2}
\eqalign{
& R_x=D_x^2-\frac{2u_xu_{xx}}{u_x^2+1}D_x-\left((u_x^2+1)+\frac{u_x^2u_{xx}^2}{(u_x^2+1)^2}\right)+\\
& \qquad +u_xD_x^{-1}\left(u_{xx}+\frac{(3u_x^2+1)u_{xx}u_{xxx}}{u_x(u_x^2+1)^2}-\frac{(3u_x^2-1)u_{xx}^3}{(u_x^2+1)^3}\right),\\
& R_y=D_y^2-\frac{2u_yu_{yy}}{u_y^2+1}D_y-\left((u_y^2+1)+\frac{u_y^2u_{yy}^2}{(u_y^2+1)^2}\right)+\\
& \qquad +u_yD_y^{-1}\left(u_{yy}+\frac{(3u_y^2+1)u_{yy}u_{yyy}}{u_y(u_y^2+1)^2}-\frac{(3u_y^2-1)u_{yy}^3}{(u_y^2+1)^3}\right).}
\end{eqnarray}

{\bf Proposition 3.} Operators $R_x$ and $R_y$ given in (\ref{R-2}) define recursion operators generating hierarchies of the symmetries in the directions of $x$ and respectively $y$.

For example by applying these operators to the generators of classical symmetries $u_x$ and $u_y$ we get 
\begin{eqnarray*}
& R_x(u_x)=u_{xxx}-\frac{3u_xu^2_{xx}}{2(u^2_x+1)}-\frac{1}{2}u_x^3,\\
& R_y(u_y)=u_{yyy}-\frac{3u_yu^2_{yy}}{2(u^2_y+1)}-\frac{1}{2}u_y^3.
\end{eqnarray*}
Thus the first members of the hierarchies are: 
$$u_t=u_{xxx}-\frac{3u_xu^2_{xx}}{2(u^2_x+1)}-\frac{1}{2}u_x^3, \qquad u_\tau=u_{yyy}-\frac{3u_yu^2_{yy}}{2(u^2_y+1)}-\frac{1}{2}u_y^3.$$

\section{Construction of the Lax pair of the usual form}

In this section we show that the third order scalar differential equations (\ref{3.5-oim}), (\ref{3.1-oim}), (\ref{oim2-x}), (\ref{oim2-y}) derived in the previous sections  from $3\times 3$ Lax pairs (\ref{1.28}), (\ref{1.29}) and (\ref{1.28}), (\ref{2.17}) can be used for constructing usual $2\times 2$ Lax pairs for the equations (\ref{1.1}) and (\ref{2.1}).

\subsection{Derivation of the Lax pairs for the first equation}

Let us derive the Lax pair for the equation 

\begin{equation}\label{1.1D}
u_{xy}=\sqrt{u_x}\sqrt{u_y^2+1}.
\end{equation}
To do this, we first lower the order of the third-order equation (\ref{3.5-oim}), rewritten for the sake of simplicity as 
\begin{equation}\label{3.5-oimD}
v_{xxx}-\frac{u_{xx}}{u_x}v_{xx}-\left(u_x+\frac{u_{xxx}}{2u_x}-\frac{3u_{xx}^2}{4u_x^2}+\frac{1}{4\lambda}\right)v_{x}-\frac{u_{xx}}{2}v=0
\end{equation} 
by using the following trick. We look for a second order ODE with the sought function~$v$
\begin{equation}\label{3.5-oimDI}
v_{xx}=F(v_{x},v,u_{xx}, u_x, u)
\end{equation} 
generally nonlinear, that would be consistent with  (\ref{3.5-oimD}) for all values of the dynamical variables $u$, $u_x$, $u_{xx}, ...$ of the equation (\ref{1.1D}). Then as it is easily seen the desired function $F$ should satisfy the next condition
\begin{equation*}
D_x(F)=\frac{u_{xx}}{u_{x}}F+\left(u_x+\frac{1}{2}\frac{u_{xxx}}{u_{x}}-\frac{3u_{xx}^2}{4u_x^2}+\frac{1}{4\lambda}\right)v_x+\frac{1}{2}u_{xx}v
\end{equation*} 
or, in the enlarged form:
\begin{eqnarray}\label{oim-cond}
\eqalign{
u_x^2F_{v_x}F+u_x^2v_xF_{v}+u_x^3F_{u}+u_x^2u_{xx}F_{u_x}+u_x^2u_{xxx}F_{u_{xx}}-u_xu_{xx}F-\cr
\qquad -\frac{1}{2} u_x^2u_{xx}v- u_x^3v_x-\frac{1}{2} u_xu_{xxx}v_x+\frac{3}{4} u_{xx}^2v_x-\frac{1}{4\lambda}u_x^2v_x=0.
}
\end{eqnarray} 
We collect the coefficients in front of the variable $u_{xxx}$ that is considered as independent one and get an equation
\begin{equation*}
u_xF_{u_{xx}}-\frac{1}{2} v_x=0,
\end{equation*}
which implies:
\begin{equation*}
F(v_{x},v,u_{xx}, u_x, u)=\frac{1}{2}\frac{u_{xx}}{u_{x}}v_x+F_1(v_{x},v,u_x,u).
\end{equation*}
Let's substitute the obtained expression into (\ref{oim-cond}) and compare the coefficients of $u_{xx}$, as a result we find
\begin{equation*}
F_1(v_{x},v,u_x,u)=u_xv+F_2\left(v,u,\frac{u_{x}}{v_x^2}\right)v_x.
\end{equation*}
Therefore function $F$ is represented as
\begin{equation*}
F(v_{x},v,u_{xx}, u_x, u)=\frac{1}{2}\frac{u_{xx}}{u_{x}}v_x+u_xv+F_2\left(v,u,\frac{u_{x}}{v_x^2}\right)v_x.
\end{equation*}
Now we simplify equation (\ref{oim-cond}) by virtue of the found representation. First we make a change of the variables $u_x=hv_x^2$, i.e. now $h$ is considered as a new functional parameter:
\begin{eqnarray}\label{oim-cond-h}
\fl (F_2)'_uhv_x^2-\left(2(F_2)'_hvh^2-F_2hv-(F_2)'_v\right)v_x-\frac{1}{4\lambda}\left(8\lambda hF_2(F_2)'_h-4\lambda (F_2)^2+1\right)=0.
\end{eqnarray}
Note that equation (\ref{oim-cond-h}) splits into three equations, since the variable $u_x$ (and hence variable $h$) is considered independent:
\begin{eqnarray*}
1) \quad (F_2)'_u=0,\\
2) \quad 2(F_2)'_hvh^2-F_2hv-(F_2)'_v=0,\\
3) \quad 8\lambda hF_2(F_2)'_h-4\lambda (F_2)^2+1=0.
\end{eqnarray*}
Obviously the first equation implies that $F_2$ doesn't depend on $u$. Further by solving the third equation we find
\begin{eqnarray}\label{F2}
F_2(v,h)=-\frac{1}{2\sqrt{\lambda}}\sqrt{4\lambda hF_3(v)+1}.
\end{eqnarray}
In the next step we focus on the second equation, that is also simplified due to (\ref{F2}) and gives rise to
\begin{eqnarray*}
F_3(v)=-\frac{v^2}{4\lambda}+C,
\end{eqnarray*}
where $C$ is an arbitrary constant. Finally, by summarizing all of the reasoning above and returning back from the variable $h$ to the variable $u_x$, we arrive at the following result
\begin{equation*}
F(v_{x},v,u_{xx}, u_x, u)=\frac{1}{2}\frac{u_{xx}}{u_{x}}v_x+u_xv-\frac{1}{2\sqrt{\lambda}}\sqrt{v_x^2-(v^2-4\lambda C)u_x}.
\end{equation*}
Thus, the desired equation (\ref{3.5-oimDI}) should look like
\begin{equation}\label{vxx-final}
v_{xx}=\frac{1}{2}\frac{u_{xx}}{u_{x}}v_x+u_xv-\frac{1}{2\sqrt{\lambda}}\sqrt{v_x^2-u_x(v^2-4\lambda C)}.
\end{equation}
To find the Lax pair in addition to (\ref{vxx-final}), we also use the equation
\begin{equation}\label{vy-final}
v_{y}=-\sqrt{\lambda}\frac{\sqrt{u_y^2+1}}{\sqrt{u_x}}\sqrt{v_x^2-u_x(v^2-4\lambda C)},
\end{equation}
which is a differential consequence of the equation (\ref{vxx-final}). Indeed, the equation (\ref{vy-final}) is obtained by applying operator $D_y$ to (\ref{vxx-final}) and further simplifying with the equation (\ref{1.1D}) and its linearization
\begin{equation}\label{lin1}
v_{xy}=\frac{\sqrt{u_y^2+1}}{2\sqrt{u_x}}v_x+\frac{u_y\sqrt{u_x}}{\sqrt{u_y^2+1}}v_y.
\end{equation}
In formulas (\ref{vxx-final})-(\ref{lin1}) we assume that the constant of integration vanishes: $C=0$ and make a change of the variables:
\begin{eqnarray}
v=2\varphi\psi,  \label{v}\\
v_x=\sqrt{u_x}(\varphi^2+\psi^2). \label{vx}
\end{eqnarray}
The compatibility conditions of the equations (\ref{v}) and (\ref{vx}) and respectively equations (\ref{vx}) and (\ref{vxx-final}) imply 
\begin{eqnarray*}
2\varphi_x\psi+2\varphi\psi_x-\sqrt{u_x}(\varphi^2+\psi^2)=0,  \\
(2\varphi\varphi_x+2\psi\psi_x)-2\sqrt{u_x}\varphi\psi-\frac{1}{2\sqrt{\lambda}}(\varphi^2-\psi^2)=0.  
\end{eqnarray*}
By combining these two equations one can obtain the following system of linear equations, defining the $x$-part of the Lax pair to (\ref{1.1}),
\begin{equation}\label{Lax-x-1}
\left \{
\begin{array}{l}
  \displaystyle
\varphi_x=\xi\varphi+\frac{1}{2}\sqrt{u_x}\psi, \\[1ex]
  \displaystyle
\psi_x=\frac{1}{2}\sqrt{u_x}\varphi-\xi\psi,  
\end{array} 
\right.
\end{equation}
where $\xi=\frac{1}{4\sqrt{\lambda}}$. Now it remains to derive the $y$-part of the Lax pair. To this end we apply $D_y$ to equations (\ref{v}) and (\ref{vx}) and then simplify in virtue of (\ref{vy-final}), (\ref{lin1}). As a result we get 
\begin{equation}\label{Lax-y-1}
\left \{
\begin{array}{l}
  \displaystyle
\varphi_y=\frac{1}{8}\xi^{-1}\left(\sqrt{u_y}\varphi-\sqrt{u_y^2+1}\psi\right), \\[1ex]
  \displaystyle
\psi_y=\frac{1}{8}\xi^{-1}\left(\sqrt{u_y^2+1}\varphi-\sqrt{u_y}\psi\right).  
\end{array} 
\right.
\end{equation}
It is easily verified by a straightforward computation that a pair of systems (\ref{Lax-x-1}),  (\ref{Lax-y-1}) determine the Lax pair for equation (\ref{1.1}).

{\bf Remark 1.} The triple of equations (\ref{vxx-final}), (\ref{vy-final}), (\ref{lin1}) are here used for deriving the usual Lax pair for the nonlinear equation (\ref{1.1}). However these equations are important by themselves. Note that a system of these three equations is consistent if and only if $u=u(x,y)$ is a solution to (\ref{1.1}). Due to this property the triple is very convenient basis to deriving Dubrovin type systems, describing algebro-geometric solutions of equation (\ref{1.1}) (see, \cite{HabKhaJPA20}, \cite{UMJ21}, \cite{Non23}).

\subsection{Construction of the Lax pair for the second equation}

Now we concentrate on equation
\begin{equation}\label{2.1D}
u_{xy}=\sqrt{u_x^2+1}\sqrt{u_y^2+1}.
\end{equation}
Since the two equations (\ref{1.1D}) and (\ref{2.1D}) are very close to each other we explain the procedure of derivation of the Lax pair in a very short way. 

We begin with the third order ordinary differential equation (\ref{oim2-y}). At first we reduce its order and find 
\begin{equation}\label{4.1}
\fl \qquad v_{yy}=\frac{u_yu_{yy}}{u_y^2+1}v_y+u_y\sqrt{\lambda v_y^2+(u_y^2+1)(\lambda^2 v^2-\lambda v^2+C)}-(\lambda-1)(u_y^2+1)v.
\end{equation} 
Then we derive a nonlinear first order PDE 
\begin{equation}\label{4.2}
v_{x}=\frac{\sqrt{u_x^2+1}}{\lambda\sqrt{u_y^2+1}}\sqrt{\lambda v_y^2+(u_y^2+1)(\lambda^2v^2-\lambda v^2+C)},
\end{equation} 
by applying operator $D_x$ to (\ref{4.1}) and simplifying due to equation (\ref{2.1}) and its linearization 
\begin{equation}\label{lin2}
v_{xy}=\frac{u_x\sqrt{u_y^2+1}}{2\sqrt{u_x^2+1}}v_x+\frac{u_y\sqrt{u_x^2+1}}{\sqrt{u_y^2+1}}v_y.
\end{equation}
We take $C=0$ and pass to new variables $\varphi$, $\psi$ in the system of equations (\ref{4.1}), (\ref{4.2}), (\ref{lin2}) by setting 
\begin{eqnarray}
v=2\varphi\psi,  \label{v-2}\\
v_y=\sqrt{\lambda-1}\sqrt{u_y^2+1}(\varphi^2-\psi^2). \label{vy-2}
\end{eqnarray}
The compatibility condition of the equations (\ref{v-2}) and (\ref{vy-2}) as well as the consistency of (\ref{vy-2}) and (\ref{4.1}) imply the $y$-part of the Lax pair to equation (\ref{2.1})
\begin{equation}\label{4.5}
\left \{
\begin{array}{l}
  \displaystyle
\varphi_y=\frac{\sqrt{\lambda}u_y}{2}\varphi-\frac{\sqrt{\lambda-1}\sqrt{u_y^2+1}}{2}\psi,\\[1ex]
  \displaystyle
\psi_y=\frac{\sqrt{\lambda-1}\sqrt{u_y^2+1}}{2}\varphi-\frac{\sqrt{\lambda}u_y}{2}\psi.
\end{array} 
\right.
\end{equation}
Let us derive the $x$-part of the Lax pair. To this end we apply operator $D_x$ on (\ref{v-2}) and (\ref{vy-2}) and then simplify by virtue of (\ref{4.2}) and (\ref{lin2}). As a result we obtain 
\begin{equation}\label{4.6}
\left \{
\begin{array}{l}
  \displaystyle
\varphi_x=\frac{u_x}{2\sqrt{\lambda}}\varphi+\frac{\sqrt{\lambda-1}\sqrt{u_x^2+1}}{2\sqrt{\lambda}}\psi,\\[1ex]
  \displaystyle
\psi_x=\frac{\sqrt{\lambda-1}\sqrt{u_x^2+1}}{2\sqrt{\lambda}}\varphi-\frac{u_x}{2\sqrt{\lambda}}\psi.
\end{array} 
\right.
\end{equation}
The obtained Lax pair (\ref{4.5}), (\ref{4.6}) depends on the spectral parameter defined on an elliptic curve. But the situation is easily improved by taking a new parameter $\xi$ defined as follows $4\lambda=\left(\xi+\xi^{-1}\right)^2$:
\begin{equation*}
\left \{
\begin{array}{l} \displaystyle
\varphi_y=\frac{\xi+\xi^{-1}}{4}u_y\varphi-\frac{\xi-\xi^{-1}}{4}\sqrt{u_y^2+1}\psi,\\[1ex]
  \displaystyle
\psi_y=\frac{\xi-\xi^{-1}}{4}\sqrt{u_y^2+1}\varphi-\frac{\xi+\xi^{-1}}{4}u_y\psi,
\end{array} 
\right.
\end{equation*}
\begin{equation*}
\left \{
\begin{array}{l}
  \displaystyle
\varphi_x=\frac{1}{\xi+\xi^{-1}}u_x\varphi+\frac{\xi-\xi^{-1}}{2(\xi+\xi^{-1})}\sqrt{u_x^2+1}\psi,\\[1ex]
  \displaystyle
\psi_x=\frac{\xi-\xi^{-1}}{2(\xi+\xi^{-1})}\sqrt{u_x^2+1}\varphi-\frac{1}{\xi+\xi^{-1}}u_x\psi.
\end{array} 
\right.
\end{equation*}

\section*{Conclusions}

The class of nonlinear integrable equations of hyperbolic type of the form (\ref{Iy.1}) has been actively studied over the past few centuries. With the discovery of the inverse scattering transform method, interest in equations of this class increased significantly (see, for example, \cite{Novikovbook}). Of particular interest is the most important problem of enumerating all integrable equations (\ref{Iy.1}) of the soliton type, i.e., equations without non-trivial characteristic integrals. One of the main tools for classifying integrable equations is the symmetry approach. Note that the first classification result within this approach was obtained in \cite{ZhSh79}. The current state of the problem of classifying integrable partial differential equations, as well as references to publications on this topic, can be found in the book \cite{Sokolov}. As noted in \cite{Sokolov}, at present the problem of a complete description of integrable equations of the form (\ref{Iy.1}) remains an open problem.

Liouville-type equations constitute an important subclass of integrable equations of the form (\ref{Iy.1}). They are characterized by the fact that the sequence of Laplace invariants $h_{[i]}$ ends with zero in both directions (\cite{Anderson}, \cite{Zhiber}). This condition was used in \cite{Zhiber} as a classification criterion for Liouville-type equations.

Note that for soliton equations of the form (\ref{Iy.1}), i.e. for equations of the sine-Gordon type, all Laplace invariants are nonzero, and the Laplace sequence does not terminate on either side. Our conjecture is that this subclass of integrable equations is completely characterized by the fact that the Laplace sequence for them admits a finite reduction of the form (\ref{Iy.12}). The conjecture is confirmed by several examples from work \cite{HabKhaPo} and examples (\ref{1.1}), (\ref{2.1}) above. We assume that the condition for the presence of such a reduction of the Laplace sequence can be chosen as a classification criterion in the problem of enumerating integrable soliton equations of the form (\ref{Iy.1}).


\section*{References}

\begin{thebibliography}{20}

\bibitem{Laplace} Laplace P S 1776 {\it Recherches sur le calcul int\'egral aux diff\'erences partielles} (M\'emoires de l'acad\'emie royale de Sciences de Paris)

\bibitem{Liouville} Liouville J  1853 Sur l'\'equation aux diff\'erences partielles ${d^2\log \lambda \over du dv}\pm {\lambda \over 2a^2}=0$ {\it J. Math. Pures Appl.} {\bf 18} 71-72

\bibitem{Anderson} Anderson I M and Kamran N 1997 The variational bicomplex for second order scalar partial differential equations in the plane  {\it Duke Math. J.} {\bf 87} 265-319

\bibitem{Ferapontov} Ferapontov E V 1997 Laplace transformations of hydrodynamic type systems in Riemann invariants: periodic sequences {\it J. Phys. A: Math. Gen.}  {\bf 30} 6861-6878  

\bibitem{Dynnikov} Novikov S P and Dynnikov I A 1997 Discrete spectral symmetries of low-dimensional differential operators and difference operators on regular lattices and two-dimensional manifolds {\it Russian Math. Surveys} {\bf 52} 1057-1116

\bibitem{Adler} Adler V E and Startsev S Ya 1999 Discrete analogues of the Liouville equation {\it Theoret. and Math. Phys.} {\bf 121} 1484-1495

\bibitem{Zhiber} Zhiber A V and Sokolov V V 2001 Exactly integrable hyperbolic equations of Liouville type {\it Russ. Math. Surv.} {\bf 56} 61-101

\bibitem{Ganzha} Ganzha E I and Tsarev S P 2001 {\it Integration of classical series $A_n$, $B_n$, $C_n$ of exponential systems} (Krasnoyarsk: Krasnoyarsk State Pedagogical University Press)

\bibitem{HabKhaPo}  Habibullin I T, Khakimova A R and Poptsova M N 2016 On a method for constructing the Lax pairs for nonlinear integrable equations {\it J. Phys. A: Math. Theor.} {\bf 49} 035202 

\bibitem{DubrovinMatveevNovikov} Dubrovin B A, Matveev V B and Novikov S P 1976 Non-linear equations of Korteweg-de Vries type, finite-zone linear operators, and Abelian varieties {\it Russ. Math. Surv.} {\bf 31} 59-146 

\bibitem{HabKhaJPA17}	Habibullin I T and Khakimova A R 2017 On a method for constructing the Lax pairs for integrable models via a quadratic ansatz {\it J. Phys. A: Math. Theor.} {\bf 50} 305206

\bibitem{HabKhaJMS21} Habibullin I T and Khakimova A R 2020 Invariant manifolds of hyperbolic integrable equations and their applications {\it J. Math. Sci.} {\bf 257} 410-423

\bibitem{SokMesh} Meshkov A G and Sokolov V V 2011 Hyperbolic equations with third-order symmetries {\it Theoret. and Math. Phys.} {\bf 166}  43-57 

\bibitem{HabKhaJPA20} Habibullin I T and Khakimova A R 2020 Invariant manifolds and separation of the variables for integrable chains {\it J. Phys. A: Math. Theor.} {\bf 53} 385202 

\bibitem{UMJ21} Habibullin I T, Khakimova A R and Smirnov A O 2021 Generalized invariant manifolds for integrable equations and their applications {\it Ufa Math. J.} {\bf 13} 135-151

\bibitem{Non23} Habibullin I T, Khakimova A R and Smirnov A O 2023 Construction of exact solutions to the Ruijsenaars-Toda lattice via generalized invariant manifolds {\it Nonlinearity} {\bf 36} 231-254

\bibitem{Novikovbook} Novikov S, Manakov S V, Pitaevskii L P and Zakharov V E 1984 {\it Theory of Solitons: The Inverse Scattering Method} (Springer Science, Business Media)

\bibitem{ZhSh79} Zhiber A V and Shabat A B 1979 The Klein–Gordon equation with nontrivial group {\it Dokl. Akad. Nauk SSSR} {\bf 247} 1103-1107

\bibitem{Sokolov} Sokolov V 2020 {\it Algebraic structures in Integrability} (World Scientific)

\end{thebibliography}
\end{document}